\documentclass[aps,twocolumn,prd,preprintnumbers, 10pt]{revtex4-1}
\pdfoutput=1
\usepackage[hidelinks]{hyperref}
\usepackage{amssymb}
\usepackage{amsfonts}
\usepackage{graphicx}
\usepackage{epstopdf}
\usepackage{dcolumn}
\usepackage{amsmath}
\usepackage{latexsym,bm}
\usepackage{amsthm}
\usepackage{slashed}
\usepackage{float}
\usepackage{color}
\usepackage{url}
\usepackage{longtable}
\usepackage{rotating}
\usepackage[normalem]{ulem}
\usepackage{faktor}
\usepackage{tikz-cd}
\usepackage{tensor}
\usepackage{array}

\newcommand{\bbeta}{\beta}

\newcommand{\cC}{\mathcal{C}}
\newcommand{\cD}{\mathcal{D}}

\newcommand{\cF}{\mathcal{F}}
\newcommand{\cG}{\mathcal{G}}

\newcommand{\cK}{\mathcal{K}}
\newcommand{\cL}{\mathcal{L}}

\newcommand{\cN}{\mathcal{N}}
\newcommand{\cO}{\mathcal{O}}

\newcommand{\cR}{\mathcal{R}}

\newcommand{\cV}{\mathcal{V}}

\renewcommand{\d}{\text{d}}

\newcommand{\nn}{\nonumber}

\newcommand{\be}{\begin{equation}}
\newcommand{\ee}{\end{equation}}
\newcommand{\ba}{\begin{aligned}}
\newcommand{\ea}{\end{aligned}}

\newcommand{\bea}{\begin{eqnarray}}
\newcommand{\eea}{\end{eqnarray}}

\def\unit{{1\kern-.65ex {\rm l}}}
\def\1{{1\kern-.65ex {\rm l}}}

\newcommand{\beq}{\begin{equation}}
\newcommand{\eeq}{\end{equation}}

\usepackage{verbatim}
\usepackage{tikz}
\usetikzlibrary{arrows,snakes,shapes.arrows,decorations.markings}
     \tikzset{>=triangle 90}
     \tikzstyle{gr}=[draw,circle,green!50!black,fill=green!50!black,scale=.6]
     \tikzstyle{Bl}=[draw,circle,blue,scale=.7]
     \tikzstyle{R}=[draw,circle,fill=red,scale=.7]
     \tikzstyle{bl}=[draw,circle,fill=black,scale=.2]
     \tikzstyle{bbc}=[draw,circle,fill=black,scale=.75]
     \tikzstyle{bbcs}=[draw,circle,fill=black,scale=.5]
     \tikzstyle{rc}=[circle,fill=red,scale=.6]
     \tikzstyle{wc}=[draw,circle,scale=.75]
\usepackage{array} 
\usepackage{multirow} 
\usepackage{colortbl} 
\usepackage{stfloats}  
\usepackage{cellspace}
\usepackage{xcolor}   

\newcommand{\xdasharrow}[2][->]{
\tikz[baseline=-\the\dimexpr\fontdimen22\textfont2\relax]{
\node[anchor=south,font=\scriptsize, inner ysep=1.5pt,outer xsep=2.2pt](x){#2};
\draw[shorten <=3.4pt,shorten >=3.4pt,dashed,#1](x.south west)--(x.south east);
}
}

\def\bar{\overline}

\def\^{\wedge}

\def\cC{{\mathcal C}}

\def\cD{{\mathcal D}}

\def\cK{{\mathcal K}}

\def\cN{{\mathcal N}}
\def\cO{{\mathcal O}}

\def\cR{{\mathcal R}}

\newcount\hour \newcount\minute
\hour=\time \divide \hour by 60
\minute=\time
\def\now{%
\ifnum \hour<13
  \ifnum \hour=0 \advance \hour by 12 \number\hour:\else \number\hour:\fi%
     \ifnum \minute<10 0\fi%
     \number\minute%
\ A.M.%
\else \advance \hour by -12 \number\hour:%
  \ifnum \minute<10 0\fi%
  \number\minute%
  \ P.M.%
\fi%
}

\usepackage{tikz}
\usetikzlibrary{arrows}
\usetikzlibrary{shapes.geometric,calc,arrows, positioning,shapes.misc,decorations.markings}
\tikzset{
  big arrow/.style={
    decoration={markings,mark=at position 1 with {\arrow[scale=2,#1]{>}}},
    postaction={decorate},
    shorten >=0.4pt},
  big arrow/.default=black}
\usetikzlibrary{external}
\usetikzlibrary{positioning}
\usetikzlibrary{calc}
\tikzset{gauge-node/.style={shape=circle, draw, minimum width=.6cm}}

\pgfdeclarelayer{edgelayer}
\pgfdeclarelayer{nodelayer}
\pgfsetlayers{edgelayer,nodelayer,main} 
\tikzstyle{none}=[inner sep=0pt] 

\tikzstyle{NodeCross}=[draw, shape=circle, cross out, inner sep=0pt, minimum size=6pt,line width=0.25mm]
\tikzstyle{Circle}=[draw, shape=circle, black, inner sep=0pt, minimum size=6pt]
\tikzstyle{rtriangle}=[fill=black, regular polygon, regular polygon sides=3, rotate=90, inner sep=0pt, minimum size=8pt]
\tikzstyle{ltriangle}=[fill=black, regular polygon, regular polygon sides=3, rotate=270, inner sep=0pt, minimum size=8pt]
\tikzstyle{rtriangleblue}=[fill={rgb,255: red,17; green,160; blue,255}, regular polygon, regular polygon sides=3, rotate=90, inner sep=0pt, minimum size=8pt]
\tikzstyle{ltriangleblue}=[fill={rgb,255: red,17; green,160; blue,255}, regular polygon, regular polygon sides=3, rotate=270, inner sep=0pt, minimum size=8pt]
\tikzstyle{rtrianglegreen}=[fill={rgb,255: red,69; green,255; blue,28}, regular polygon, regular polygon sides=3, rotate=90, inner sep=0pt, minimum size=8pt]
\tikzstyle{ltrianglegreen}=[fill={rgb,255: red,69; green,255; blue,28}, regular polygon, regular polygon sides=3, rotate=270, inner sep=0pt, minimum size=8pt]
\tikzstyle{Uprtriangle}=[fill=black, regular polygon, regular polygon sides=3, rotate=0, inner sep=0pt, minimum size=8pt]
\tikzstyle{Downltriangle}=[fill=black, regular polygon, regular polygon sides=3, rotate=180, inner sep=0pt, minimum size=8pt]
\tikzstyle{rtriangleAmber}=[fill={rgb,255: red, 191; green, 144; blue, 63}, regular polygon, regular polygon sides=3, rotate=90, inner sep=0pt, minimum size=8pt]
\tikzstyle{UprtriangleViolett}=[fill={rgb,255: red,255; green,0; blue,0}, regular polygon, regular polygon sides=3, rotate=0, inner sep=0pt, minimum size=8pt]

\tikzstyle{Downltriangle}=[fill=black, regular polygon, regular polygon sides=3, rotate=180, inner sep=0pt, minimum size=8pt]
\tikzstyle{UpRighttriangle}=[fill=black, regular polygon, regular polygon sides=3, rotate=45, inner sep=0pt, minimum size=8pt]
\tikzstyle{UpLefttriangle}=[fill=black, regular polygon, regular polygon sides=3, rotate=315, inner sep=0pt, minimum size=8pt]
\tikzstyle{DownRighttriangle}=[fill=black, regular polygon, regular polygon sides=3, rotate=135, inner sep=0pt, minimum size=8pt]
\tikzstyle{DownLighttriangle}=[fill=black, regular polygon, regular polygon sides=3, rotate=225, inner sep=0pt, minimum size=8pt]

\tikzstyle{Star}=[draw, shape=star, fill=black, star points=8, inner sep=0pt, minimum size=8pt]

\tikzstyle{DashedLine}=[-, densely dashed, line width=0.25mm]
\tikzstyle{DashedLineBrown}=[-, densely dashed, line width=0.25mm, draw={rgb,255: red,155; green,103; blue,51}]
\tikzstyle{DashedLineFall}=[-, densely dashed, line width=0.25mm, draw={rgb,255: red,195; green,0; blue,0}]
\tikzstyle{DashedLineViolett}=[-, densely dashed, line width=0.25mm, draw={rgb,255: red,139; green,41; blue,148}]
\tikzstyle{DottedLine}=[-, dotted, line width=0.25mm]
\tikzstyle{BlueLine}=[-, fill=none, draw={rgb,255: red,17; green,160; blue,255}, line width=0.25mm]
\tikzstyle{GreenLine}=[-, fill=none, draw={rgb,255: red,69; green,255; blue,28}, line width=0.25mm]
\tikzstyle{RedLine}=[-, draw={rgb,255: red,191; green,0; blue,0}, fill=none, line width=0.25mm]
\tikzstyle{DashedLineRed}=[-, densely dashed, fill=none, draw={rgb,255: red,191; green,0; blue,0}, line width=0.25mm]
\tikzstyle{ThickLine}=[-, line width=0.25mm]
\tikzstyle{ViolettLine}=[-, draw={rgb,255: red,132; green,60; blue,191}, fill=none, line width=0.25mm]
\tikzstyle{ViolettDashedLine}=[-, densely dashed, draw={rgb,255: red,132; green,60; blue,191}, fill=none, line width=0.25mm]
\tikzstyle{AmberLine}=[-, draw={rgb,255: red,191; green,144; blue,63}, fill=none, line width=0.25mm]
\tikzstyle{DashedRedThick}=[-, densely dashed, fill=none, draw={rgb,255: red,191; green,0; blue,0}, line width=0.40mm]
\tikzstyle{DashedBlueThick}=[-, densely dashed, fill=none, black, line width=0.40mm]

\makeatother

\begin{document}

\title{Non-Invertible Symmetries from Holography and Branes}

\author{Fabio Apruzzi$^{1}$}
\author{Ibrahima Bah$^2$}
\author{Federico Bonetti$^3$}
\author{Sakura Sch\"afer-Nameki$^3$}

\affiliation{$^{1}$ AEC for Fundamental Physics, ITP, 
University of Bern, Sidlerstrasse 5, 3012 Bern, Switzerland}
\affiliation{$^{2}$ Department of Physics and Astronomy, Johns Hopkins University, Baltimore, MD 21218, USA}
\affiliation{$^3$Mathematical Institute, University
of Oxford, Woodstock Road, Oxford, OX2 6GG, United Kingdom}


\begin{abstract}
\noindent 
We propose a systematic approach to deriving symmetry generators of Quantum Field Theories in holography. 
Central to this analysis are the Gauss law constraints in
the Hamiltonian quantization of Symmetry Topological Field Theories (SymTFTs), which are obtained from supergravity. 
In turn, we  realize the symmetry generators from world-volume theories of D-branes in holography.
Our main focus is on non-invertible symmetries, which have emerged in the past year as a new type of symmetry in $d\geq 4$ QFTs. 
We exemplify our proposal in the holographic confinement setup, dual to 4d $\mathcal{N}=1$ Super-Yang Mills. In the brane-picture, the fusion of non-invertible symmetries naturally arises from the Myers effect on D-branes. In turn, their action on line defects is modeled by the Hanany-Witten effect. 
\end{abstract}




\maketitle


\noindent{\bf Introduction.}
The study of quantum dynamics is at the heart of uncovering any fundamental principles of nature. From various points of view, in condensed matter physics, mathematical physics and quantum field theory, such explorations have established  the study of symmetries as an essential ``backbone" of quantum systems.  
It  thus comes as a genuine surprise in the past year where a dramatic extension to symmetries in 4d quantum field theories (QFTs) were uncovered, which unlike ordinary ones that form  groups, obey fusion-like composition laws. 
These {\it non-invertible symmetries}  are well-established in $d=2,3$, however, they are unexpected in $d\geq 4$. Within the past year various systematic approaches to the construction of non-invertible symmetries have appeared in \cite{Kaidi:2021xfk, Choi:2021kmx, Bhardwaj:2022yxj, Choi:2022zal, Bhardwaj:2022lsg, Lin:2022xod, Bartsch:2022mpm}. Physical implications include characterization of de/confining vacua and constraints on pion decays \cite{Choi:2022jqy, Cordova:2022ieu}, and other applications appeared in \cite{Heidenreich:2021xpr, Choi:2022jqy, Cordova:2022ieu, Choi:2022rfe, Bashmakov:2022jtl, Kaidi:2022uux,  Damia:2022bcd}.

All constructions thus far rely on field theory methods. Here we provide the  holographic perspective from symmetry inflow,  supergravity and branes. Other preliminary aspects of holography and non-invertible symmetries have been recently studied in \cite{Damia:2022bcd, Damia:2022rxw, Benini:2022hzx}.
  Fundamental  for the holographic construction is the {\it Symmetry Topological Field Theory (SymTFT)} \cite{Freed:2012bs, Gaiotto:2020iye, Apruzzi:2021nmk,Apruzzi:2022dlm}, which naturally arises in brane/holographic setups from the anomaly polynomial and inflow \cite{Harvey:1998bx,Freed:1998tg,Bah:2019rgq, Bah:2020jas,Bah:2020uev}. 
The SymTFT on $W_{d+1}$ encodes the full symmetry structure -- the background fields for global symmetries and  their 't Hooft anomalies -- of a QFT on $W_d = \partial W_{d+1}$.
When placed on a slab with boundaries $W_d$ and $M_d$ and gapped boundary condition on $M_d$, the SymTFT reduces to the anomaly theory of the QFT.

\begin{figure}
\centering
$
\begin{tikzpicture}
\draw [thick, red](0,0) -- (0,2);
\draw [thick,blue](1,0) -- (1,2);
\node[red] at (0, -0.5) {D3};
\node[blue] at (1,-0.5) {D5};
\node[black] at (2.5, 0.6) {$\Longrightarrow$} ;
\begin{scope}[shift={(4,0)}]
\draw [thick,blue](0,0) -- (0,2);
\draw [thick,red](1,0) -- (1,2);
\draw [thick, teal] (0,1) -- (1,1);
\node[blue] at (0,-0.5) {D5};
\node[red] at (1, -0.5) {D3};
\node[teal] at (0.6, 0.7) {F1} ;
\end{scope}
\begin{scope}[shift={(0,-3.1)}]
\draw [red] (0,1) ellipse (0.1 and 0.5);
\draw [thick,blue, fill= blue, opacity= 0.2]
(1,0) -- (1,2) -- (1.5, 2.2) -- (1.5, 0.2) -- (1,0);
\node[red] at (0, -0.5) {$\mathbf{H}$};
\node[blue] at (1.2,-0.5) {${\mathcal{N}_3^{(1)}}$};
\node[black] at (2.5, 1) {$\Longrightarrow$} ;
\end{scope}
\begin{scope}[shift={(3.8,-3.1)}]
\draw [thick, teal] (0.25, 1.5) -- (1.5, 1.5);
\draw [thick, teal] (0.25, 0.5) -- (1.5, 0.5);
\draw [fill= teal, opacity=1] 
(0.25, 1.5) -- (1.5, 1.5) -- (1.5, 0.5) -- (0.25, 0.5) -- (0.25, 1.5);
\draw [thick,blue, fill= blue, opacity= 0.2]
(0,0) -- (0,2) -- (0.5, 2.2) -- (0.5, 0.2) -- (0,0);
\draw [teal, fill= teal, opacity= 1] (0.25,1) ellipse (0.1 and 0.5);
\draw [-stealth,  fill = teal, opacity= 1] (1.5,1) ellipse (0.1 and 0.5);
\draw [thick, red] (1.5,1) ellipse (0.1 and 0.5);
\node[red] at (1.5, -0.5) {$\mathbf{H}$};
\node[blue] at (0.2,-0.5) {${\mathcal{N}_3^{(1)}}$};
\end{scope}
\end{tikzpicture}
$
\caption{Top: Hanany-Witten transition, where the  $(x_0, x_3)$-plane  is displayed to show the equivalence with the field theory transition. Bottom: 't Hooft loop passing through the non-invertible defect $\mathcal{N}_3^{(1)}$ becomes attached to a topological surface operator. \label{fig:HW}}
\end{figure}
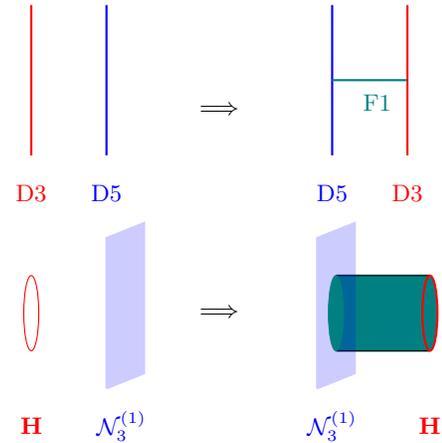


In this paper we propose a holographic derivation of the SymTFT, as well as the study of the resulting symmetries -- including non-invertible ones--  that depend on said boundary conditions. 
We derive the SymTFT by descent from the anomaly polynomial in $d+2$ dimensions, which  is encoded in the supergravity.
Motivated by the work
on BF-type theories in \cite{Belov:2004ht}, the  Hamiltonian quantization of the SymTFT on $W_{d+1}$ allows us to extract the Gauss law constraints that generate gauge symmetry transformations. Under inflow, the bulk gauge symmetry restricts to the global symmetries of the boundary theory and the bulk generators flow to the desired symmetry operators. 

This is complemented by a realization of the symmetry generators in terms of D-branes and their worldvolume theories. The bulk supergravity fields, which define the symmetries, pull back on the brane worldvolume theories. In addition, the D-branes also contribute topological sectors that dress the symmetry defect, while the kinetic terms of the brane action drop out at the boundary. These defects become non-invertible depending on the boundary conditions for the bulk fields. 
The brane setup and its dynamics towards the boundary provide a compelling holographic interpretation for the non-invertible fusion via the Myers effect  of D$p$-branes into a single D$(p+2)$-brane, which in turn implements the non-invertible fusion.

We demonstrate our proposal in the Klebanov-Strassler solution, that is dual to a flow to confining pure $\mathcal{N}=1$ $SU(M)$  Super-Yang Mills (SYM)  \cite{Klebanov:2000hb}. 
Global properties of the gauge group can be identified in holography as in \cite{Witten:1998wy} and the study of holographic confinement using the 't Hooft anomalies of higher-form symmetries was carried out in  \cite{Apruzzi:2021phx}. 
In this paper, we determine a framework to construct all symmetries in this setup, in particular the non-invertible symmetries in the  $PSU(M) = SU(M)/\mathbb{Z}_M$ theory, which map between de-/confining vacua when spontaneously broken.
In the brane-picture 
the non-invertible fusion is naturally encoded in the Myers effect on D-branes \cite{Myers:1999ps}. Furthermore, the de-/confining transition is beautifully modelled by the Hanany-Witten brane-transition \cite{Hanany:1996ie}, figure \ref{fig:HW}.
Though we focus on holographic confinement, the methods are general and can be used to study the symmetry generators of any QFT from its SymTFT.

\bigskip
\noindent{\bf Field Theory.}
Non-Invertible symmetries in QFTs in spacetime dimensions $d\geq 4$ have recently been constructed using various approaches. One is based on the presence of global symmetries that enjoy a mixed 't Hooft anomaly \cite{Kaidi:2021xfk}. For concreteness we consider a 4d gauge theory (on a spin manifold) with 0-form symmetry $\Gamma^{(0)}=\mathbb{Z}_{2M}$, whose background field is $A_1$, and 1-form symmetry $\Gamma^{(1)} =\mathbb{Z}_M$ with background field $B_2$. 
Consider the anomaly 
\be \label{field_theory_anom}
\mathcal{A} = 
-{2 \pi } { 1  \over M} \int A_1 \cup  {\mathfrak{P} (B_2)\over 2} \,, 
\ee
where $\mathfrak{P}$ is the Pontryagin square. 
This anomaly arises in 4d $\mathcal{N}=1$ supersymmetric Yang-Mills theories for instance, where $\Gamma^{(0)}$ is the chiral symmetry. There is a non-anomalous $\mathbb{Z}_2^{(0)}\subset \mathbb{Z}_{2M}^{(0)}$. 

The generalized 0- and 1-form symmetries \cite{Gaiotto:2014kfa} are generated by 3d and 2d topological defects $D_3^g(M_3)$ and $D_2^h(M_2)$, respectively, which have group composition
 $D_p^{g_1}(M_p)\otimes D_p^{g_2}(M_p) = D_p^{g_1 g_2}(M_p)$.  Due to the anomaly, the generators for $\Gamma^{(0)}$ transform non-trivially in presence of background fields for $\Gamma^{(1)}$
\be \label{eq:shiftdef}
D_3^{g}(M_3) \rightarrow D_3^g(M_3) \exp \left(\int_{M_4} -{2 \pi i \over M}  {\mathfrak{P} (B_2)\over 2} \right) \,,
\ee
where $\partial M_4 = M_3$. Gauging the 1-form symmetry makes this defect inconsistent. The proposal in \cite{Kaidi:2021xfk} is to dress the defect $D_3^{g}(M_3)$ with a minimal TQFT $\mathcal{A}^{M,p}$, which has 1-form symmetry $\mathbb{Z}_M$ and cancels the anomaly  \cite{Hsin:2018vcg}. 

For $\mathbb{Z}_M$ this is the minimal (spin) TQFT 
$\mathcal{A}^{M,1}=U(1)_{M}$.
The dressed defects are 
\be\label{Dressing}
\mathcal{N}_3^{(1)} = D_3^{(1)} \otimes \mathcal{A}^{M,1}  \,,
\ee
where the superscript labels the generator of the 0-form symmetry. 
This defect has non-invertible fusion \cite{Choi:2022zal, Choi:2022jqy}.
For $M$ odd the TQFTs obey 
$
\mathcal{A}^{M,1} \otimes \mathcal{A}^{M,1} = 
\mathcal{A}^{M,2} \otimes \mathcal{A}^{M,2}
$.
This results in the non-invertible fusion of the 3d defects in the $PSU(M)$ theory 
\be\ba  \label{eq_fusion}
\mathcal{N}_{3}^{(1)} \otimes \mathcal{N}_3^{(1)} = \mathcal{A}^{M, 2} \,  \mathcal{N}_3^{(2)} \,.
\ea\ee
Defining the conjugate $\mathcal{N}_3^{(1)\dagger} = D_3^{-1} \otimes \mathcal{A}^{M, -1}$ results in   
\be\label{NNdagger}
\mathcal{N}_3^{(1)} \otimes \mathcal{N}_3^{(1)\dagger}  
 =  \sum_{M_2 \in H_2(M_3, \mathbb{Z}_M) }{ (-1)^{Q(M_2)} D_2(M_2) \over  |H^{0}(M_3,\mathbb{Z}_M)| }\,,
\ee
which is the condensation defect of the 1-form symmetry on $M_3$ with $D_2(M_2)= e^{i2\pi\int_{M_2} b_2/M}$, where $b_2$ is the gauge field for the 1-form symmetry. 
We will now turn to supergravity/branes and show how these non-invertible symmetries are naturally implemented in this framework.

\bigskip
\noindent{\bf Symmetries from Holography.}
We illustrate the systematic approach by realizing it in the holographic confinement setup in type IIB supergravity introduced by Klebanov-Strassler
\cite{Klebanov:2000hb}.
It describes the near-horizon
geometry of  $N$ D3-branes probing
the conifold (\emph{i.e.}~the Calabi-Yau
cone over the Sasaki-Einstein 
5-manifold  $T^{1,1}\sim S^3 \times S^2$) with $M$
D5-branes on $S^2\subset T^{1,1}$. The near-horizon geometry is $W_5 \times T^{1,1}$, where the 4d space-time where the QFT lives is $W_4=\partial W_5 $, see (\ref{eq:conmet}).
We assume integral $N/M$, so that 
the duality cascade in   field
theory   ends on  
4d $\mathcal N = 1$~$SU(M)$~SYM. 
The 5d   effective action is written in terms
of $p$-form field strengths
$f_1$, $\cF_2$, $g_2$, $h_3$, $f_3$
with  Bianchi identities
\begin{align} \label{Bianchi}
df_1 = 2M \cF_2 \, , \;
dg_2 = M h_3 \, , \;
d\cF_2 = dh_3 = df_3 = 0 \ .
\end{align}
We solve the Bianchi identities \eqref{Bianchi} in terms of
\begin{align}\label{decomp}
f_1 &= f_1^{\rm b} + d c_0 + 2M A_1 \ , \;\;\;\;
\cF_2  = \cF_2^{\rm b} + dA_1 \ ,  \\
 g_2 &= g_2^{\rm b} + d\bbeta_1 + M b_2 \ , \;\; 
h_3  = h_3^{\rm b} + db_2 \ , \;\;
f_3 = f_3^{\rm b} + dc_2 \ , \nn  
\end{align}
where $A_1$, $c_0$, $b_2$, $\bbeta_1$, $c_2$
are globally defined $p$-form gauge potentials
and $f_1^{\rm b}$,
$\cF_2^{\rm b}$, $g_2^{\rm b}$, $h_3^{\rm b}$,
$f_3^{\rm b}$ are closed forms with integral
periods, representing
topologically non-trivial base-points.
From \eqref{Bianchi} it follows
that $Mh_3$ and $2M \cF_2$ are cohomologically
trivial. Assuming $W_5$ has no torsion, 
$h_3^{\rm b}= 0 = \cF_2^{\rm b}$.
The base-points $f_1^{\rm b}$,
$g_2^{\rm b}$ represent
integral lifts of classes in $H^1(W_5;\mathbb Z_{2M})$,
$H^2(W_5;\mathbb Z_M)$   describing
discrete gauge fields for
a $\mathbb Z_{2M}$ 0-form symmetry
and $\mathbb Z_M$ 1-form symmetry.

The relevant terms in the 5d bulk action 
consist of standard kinetic terms and non-trivial
topological terms. The latter 
can be extracted \cite{Apruzzi:2021phx} from the 
consistent truncation of \cite{Cassani:2010na} or  via anomaly inflow
 \cite{Bah:2020jas}, as reported in  appendix~\ref{app_bulk}.
In order to construct the symmetry generators,
it is convenient to dualize
the 0-form potential $c_0$
into a (globally defined)
3-form gauge potential $c_3$. Our task, carried out in appendix~\ref{app_bulk}, is then to write the 5d bulk action
in terms of $A_1$, $b_2$, $\bbeta_1$,
$c_2$, $c_3$ and the base-point fluxes
$f_1^{\rm b}$, $g_2^{\rm b}$,
$f_3^{\rm b}$. The final action
consists of standard kinetic terms and 
\begin{align} \label{new_top}
S_{\rm top}   = 2 \pi \int_{W_5} \Big[  
& \tfrac 12 N (b_2  dc_2 - c_2  db_2)
 + M  ( A_1  dc_3 + c_3  dA_1) \nn \\
& + N b_2  f_3^{\rm b}
+ A_1  (g_2^{\rm b} )^2 
\Big]  \ .
\end{align}

\noindent
{\bf Symmetry Generators.}
We now analyze   the 5d bulk action  in
the Hamiltonian formalism, treating the radial
direction of $W_5$ as  Euclidean time similarly to the AdS$_5$ cases
\cite{Witten:1998wy, Belov:2004ht}.
Crucially,
the action does not depend on the time
derivatives of the 
time components of the gauge potentials.
As a result, the associated canonical
momenta are identically zero.
Varying the action
with respect to the time components of 
the gauge potentials
implements the   (classical) Gauss constraints.
We denote the variation of the action
with respect to the time component of $A_1$
as $\cG_{A_1}$, and so on. We find that 
$\cG_{\bbeta_1} = \widetilde \cG_{\bbeta_1}$ and 
\begin{align} \label{Gauss}
\cG_{b_2} & = \widetilde \cG_{b_2}
- N d_4 c_2 - N   f_3^{\rm b}\,,\ 
\cG_{c_2}  =\widetilde \cG_{c_2}
+ N d_4  b_2 \\ 
\cG_{A_1} & =\widetilde \cG_{A_1}
+2M d_4  c_3
+( g_2^{\rm b})^2 \,,\ 
\cG_{c_3}  = \widetilde \cG_{c_3}
+2M d_4 A_1\,. \nonumber
\end{align}
Here, $d_4$ denotes external derivative along the spatial slice $W_4$,  
tilde the kinetic term contributions, and all fields are understood as restricted to $W_4$. The contributions $\widetilde{\cG}$ of the bulk kinetic terms
 are suppressed near the boundary \cite{Witten:1998wy, Belov:2004ht}.

We provide a detailed derivation of symmetry generators from the Gauss law constraints in appendix \ref{app_bulk}. 
For concreteness, let us illustrate the general analysis here 
by considering  
$e^{2\pi i \int_{M_4} (2M d_4  c_3
+(g_2^{\rm b})^2)}$
as it is brought to the boundary.
Our task is to define a genuine operator
on a 3-cycle $M_3$ such that,
when raised to the $2M$-th power and with
$M_3 = \partial M_4$, it reproduces
the operator $e^{2\pi i \int_{M_4} (2M d_4  c_3
+(g_2^{\rm b})^2)}$ 
from the Gauss constraint.
We consider two options.
In Option (i), we fix $g_2^{\rm b}$
at the boundary as a classical background.
This corresponds 
to 4d $\cN = 1$ $SU(M)$ SYM,
with a global electric 
$\mathbb Z_M$ 1-form symmetry
coupled to a non-dynamical
discrete 2-form field.
The genuine operator on $M_3$ in this case
is simply the standard holonomy
$e^{2\pi i \int_{M_3}  c_3}$
accompanied by  the 
$c$-number phase 
$e^{ \frac{2\pi i}{2M}\int_{M_4}(g_2^{\rm b})^2}$. This operator
obeys group-like fusion rules.
In Option (ii) we sum over $ g_2^{\rm b}$
at the boundary.
In field theory, we gauge the electric 1-form
symmetry of 4d $\cN =1$ $SU(M)$ SYM,
thereby getting the $PSU(M)$ theory.
Casting the phase 
$e^{ \frac{2\pi i}{2M}\int_{M_4}(g_2^{\rm b})^2}$
as a genuine operator on $M_3$
we can rewrite 
$\frac{1}{2M}( g_2^{\rm b})^2$ using a 3d auxiliary theory (this is a type of inflow from the bulk operator on $M_4$ to $M_3$), which we detail in appendix \ref{app_bulk}.
The symmetry generator on $M_3$ is thus
\be  \label{good_operator}
\mathcal{N}_3^{(1)}(M_3) =
\int \cD a e^{2\pi i \int_{M_3} \left( 
 c_3  + \frac 12 M a da + a   g_2^{\rm b}
\right)  } \,,
\ee 
which has the non-invertible fusion rule \eqref{eq_fusion}.

\smallskip
\noindent{\bf The far IR for $PSU(M)$.}
The $\mathbb Z^{(0)}_{2M}$ global
symmetry of 4d $\mathcal N = 1$ $SU(M)$ SYM
is spontaneously broken to $\mathbb Z^{(0)}_2$
in the far IR and the theory has $M$ confining
vacua. The 
 mixed anomaly \eqref{field_theory_anom} is matched
by a non-trivial
4d Symmetry Enhanced Topological Phase
(SET) 
\cite{Gaiotto:2014kfa}
\be  \label{4dTQFT}
{ \mathcal{L}_\text{4d}} =    M \phi dc_3 + \tfrac 12 \phi db_1 db_1
+ \Lambda_2(db_1 + Mb_2) \,,
\ee 
where $\phi$ is a compact scalar of period 1,
$c_3$, $b_2$, $b_1$ are gauge potentials and
$\Lambda_2$ a Lagrange multiplier. 
The $b_1$, $b_2$ fields are 
non-dynamical.
The possible VEVs $\langle e^{2\pi i \phi} \rangle = e^{2\pi i p/M}$ ($p=0,1,\dots,M-1$)
label the $M$ vacua,
while  
 $e^{2\pi i \int_{\cC_3} c_3}$
describes a domain wall between vacua.
The action \eqref{4dTQFT}
is invariant under the gauge transformations 
$b_1' = b_1 - M \lambda_1$,
$c_3' = c_3 + db_1 \lambda_1 - \frac 12 M \lambda_1 d\lambda_1$.
Thus $e^{2\pi i \int_{M_3} c_3}$
has a 't Hooft anomaly, consistently
with the fact that the domain walls
in the $SU(M)$ theory support
a   3d TQFT $\mathcal A^{N,-1}$ \cite{Gaiotto:2017yup}.

In \cite{Apruzzi:2021phx} it is demonstrated
how the SET \eqref{4dTQFT} emerges
from the 5d bulk couplings
in the IR geometry $T^*S^3$
(deformed conifold). In contrast to the
UV analysis above,
the IR analysis receives contributions
from both topological and kinetic terms.
The Lagrange multiplier $\Lambda_2$ is an imprint
of the St\"uckelberg pairing between $b_1$, $b_2$ in the 5d action. The scalar $\phi$
is identified as $c_0/M$.

Let us now turn to the $PSU(M)$ theory.
The far IR is still described by \eqref{4dTQFT},
but now $b_1$, $b_2$ are local dynamical fields.
Using $db_1 = -Mb_2$, we see that
the vacuum with $\langle e^{2\pi i \phi} \rangle = e^{2\pi i p/M}$ exhibits a discrete 2-form 
 gauge theory 
$\int_{M_4} \frac{pM}{2} b_2^2$.
The domain walls are no longer realized
as $e^{2\pi i \int_{M_3} c_3}$, which is not gauge invariant, but 
 precisely by (\ref{good_operator}).
Indeed, this operator raised to the $2M$-th
power with $M_3 = \partial M_4$
reduces to the manifestly gauge invariant
quantity $e^{2\pi i \int_{M_4} \left( 
2M dc_3 + db_1 db_1
\right)}$ (where $g_2^{\rm b}$ is locally modeled by $db_1$).
On the domain wall, both $a$ and $b_1$
are dynamical and summed over. 
The total 3d theory is then
an Abelian
CS theory with
levels encoded in the matrix $\left(  \begin{smallmatrix} M & 1 \\ 1 & 0
\end{smallmatrix}  \right)$. This is
a 
Dijkgraaf-Witten theory with gauge group
$\mathbb Z_1$, hence trivial, as anticipated in
\cite{Hsin:2018vcg}.

\bigskip

\noindent{\bf D-branes as Symmetry Generators.}
The topological defects also arise as boundary limits of probe branes in the bulk that are parallel to the boundary. Both in  AdS in hyper-polar coordinates and in the $W_5$ geometry of the KS solution, where the boundary sits at $r \rightarrow \infty$, the tension $T_{\rm Dp} \sim r^p $ ($p>0$), such that the DBI part of the action is subleading and the Wess-Zumino term dominates \footnote{In addition, as suggested by \cite{Witten:1998wy}, when compactifying on $S^3$ the low-energy brane action, we can consider the truncation to the (topological) sector which describes flat fields only, and ignore any irrelevant 3d kinetic terms. These will only describe the dynamics of modes with non-zero flux.}. In addition, we stress that these D5-branes are not BPS but they can be stable in the sense of \cite{Arean:2004mm} in $r \rightarrow \infty$.
The topological terms 
for a D5-brane wrapping the $S^3$ contain
the bulk forms $c_3$ (from $C_6$ on $S^3$)
and $b_1$ (from $C_4$ on $S^3$), as well as
the  $U(1)$ gauge field $a$ on the brane.
We derive the action on the defect by reducing the D5-brane Wess-Zumino action   in appendix \ref{app:D5brane}. The result reads 
\be \label{eq_D5_action}
S_\text{D5} =  2 \pi \int_{M_3 } \left( c_3 +  {M \over 2}ada  + a db_1\right) \,.
\ee
 Here $b_1$ is a local gauge field. The cohomology class of $db_1$ is identified with $g_2^{\rm b}$
and is part of the data of the $b_2$ configuration in \eqref{decomp}. It is interesting to understand what happens 
as \eqref{eq_D5_action} is pushed to the boundary. 
We always perform a path integral over $a$, which is a localized
mode on the D5-brane. We may or may not integrate over the topologically trivial part of $b_1$,
which is a bulk mode, depending on the boundary conditions.
If we do not integrate over it,
 the holonomy of $c_3$
is dressed
with the non-trivial  TQFT $U(1)_{M}$.
If we  integrate over it, it becomes a trival theory just as in the supergravity derivation. The D-branes therefore precisely give rise to the minimal TQFT stacking. 

\bigskip
\noindent {\bf Non-invertible Fusion and Myers Effect.}
To see the non-invertible fusion, we can either repeat the field theory analysis, given the explicit form of (\ref{eq_D5_action}). There is a much more elegant way to obtain the fusion directly in string theory. 
The fusion is computed by stacking two D5-branes, which gives rise to a non-abelian gauge theory. However, a non-abelian brane configuration with an orthogonal $S^2$ geometry and a non-trivial B-field undergoes the Myers effect \cite{Myers:1999ps} in reaching
the configuration with minimal energy (see appendix \ref{app:Myers}).
The end point configuration is given by a single D7-brane with two units of worldvolume gauge flux
on $S^2$. We then write $f_2 = f_2^{S^2} + da$,
with $\int_{S^2} f_2^{S^2}=2$.
From the expansion of the Wess-Zumino action of the D7, integrating on $S^2$ and $S^3$,
the terms are
\be\label{D7D5}
S_{\text{D7/2D5}} = 2\pi \int_{M_3} \left(2c_3 +M ada + 2 a db_1 \right). 
\ee
Note that this argument is applicable for any integral value of $M$.
From the brane we thus obtain the following perspective on the fusion. Each single D5-brane results in topological defects that are dressed with $U(1)_{M} = \mathcal{A}^{M, 1}$ CS theories -- thus
string theory construction  automatically results in the minimal TQFT dressing of the defects.
The ``brane-fusion" predicts the action (\ref{D7D5}), which  
is $U(1)_{2M}$ CS-theory coupled to $b_2$. 
This is obtained also by fusing two $U(1)_{M}$ theories, \footnote{
Two $U(1)_M$ theories correspond to $
\int \mathcal{D} a \mathcal{D} a'
e^{  2\pi i \int_{M_3} \left[   \frac M2  (ada + a' da' ) +(a + a')  db_1 \right]  
}$. Define $a_+ = a+a'$, so that 
integrating out sets $a_+ = 2a $, which gives 
 precisely the brane-action (\ref{D7D5}).} 
and therefore realizes the field theory fusion rule in \eqref{eq_fusion}.

It is also tempting to conjecture that the fusion of $\mathcal{N}_3^{(1)}$ with its conjugate $\mathcal{N}_3^{(1)\dagger}$ (\ref{NNdagger}) is the fusion between defects created by brane and anti-brane, with a non-trivial field configuration. This result in the condensation defect, which is the lower-dimensional brane that couples to $db_1$. This is in fact expected from tachyon condensation of the D-Dbar system \cite{Sen:1998sm}, which needs to preserve the charge under $db_1$, and thus is expected to give rise to a non-trivial condensate. This will be discussed elsewhere, and shown to correspond to a mesh of D3-branes.

\bigskip
\noindent{\bf Action on 't Hooft lines and Hanany-Witten.}
The brane perspective makes the interaction between the 't Hooft line $\mathbf{H}$ and  the non-invertible symmetry defect,  $\mathcal{N}_3^{(1)}$, manifest. 
Field-theoretically, when such a line crosses the non-invertible topological defect, a topological surface operator is created, which connects $\mathcal{N}$ and $\mathbf{H}$, see figure \ref{fig:HW}.
This effect can be derived
by observing that 
4d $\cN = 1$ $PSU(M)$ SYM
is self-dual under
$\tau^{-1}\sigma \tau^{-1}$,
where $\tau$ is stacking with 
$e^{2 \pi i \frac 1M \int \mathfrak P(B_2)/2}$,
$B_2$ is the background field for the magnetic 1-form symmetry,
and $\sigma$ is gauging 
this 1-form symmetry (cfr.~\cite{Kaidi:2021xfk} section B).
We may then realize
the non-invertible topological
defect $\mathcal N_3^{(1)}$ via
a half-space gauging argument,
and use this picture to
derive the aforementioned
action on 't Hooft lines,
along the lines of \cite{Choi:2022jqy, Choi:2022zal}.

In order to see this effect in supergravity we need to define a surface operator, which extends in the radial direction, $r$, and ends on the boundary, $\cO_2(M_2)$. The 5d bulk EOMs select a natural candidate for 
$\cO_2(M_2)$: the $b_2$ EOM imply
\be 
 -k_{f_3} d*f_3
= N f_3 + f_1 g_2
- M k_{g_2} *g_2 =: M \cF_3 \ ,
\ee 
where the $k$'s are constants from the kinetic terms, and 
$N f_3$ and $f_1 g_2 - M k_{g_2} * g_2$ are separately closed.
The latter combination encodes the 3-form field strength
of the 2-form potential dual to $b_1$.
On shell,  $\cF_3 = d\widehat a_2$ for some
globally defined 2-form potential. The operator $\cO_2(M_2)$
is identified with a Wilson surface for $\widehat a_2$,
\be
\cO_2(M_2) = e^{2\pi i \int_{M_2} \widehat a_2}, \quad \widehat a_2 = a_2 + \kappa c_2  \,,
\ee
where $\kappa = N/M$
and $a_2$ is the dual of $b_1$
(in type IIB $a_2$ comes from $C_4$ on $S^2$, and the duality is a consequence of the self-dual $F_5$ flux).

This bulk picture fits with the D-brane picture,
in which $\cO_2(M_2)$ is realized by a D1-D3-brane bound state on $S^2 \subset T^{1,1}$, or alternatively D3s with $\kappa=N/M$ units of flux supported on $S^2$ \cite{Apruzzi:2021phx}. 
In brane engineering,
a Hanany-Witten transition \cite{Hanany:1996ie}
can occur when two branes link non-trivially in spacetime
and are passed through  each other, thereby creating a new extended object
stretching between them.
In our setup this can happen 
for D3s wrapping $S^2$ and extending along the radial direction $r$ and D5s on $S^3$ localized at the boundary: 
\small
\begin{equation} \label{eq:branesyst}
    \begin{tabular}{|c|| >{\centering}p{4mm} | >{\centering}p{4mm} | >{\centering}p{4mm} | 
    >{\centering}p{4mm}
    | >{\centering}p{4mm} |
    >{\centering}p{4mm}
    |
    >{\centering}p{4mm}
    |
    >{\centering}p{4mm}
    |
    >{\centering}p{4mm}
    |
    >{\centering\arraybackslash}p{4mm}
    |}
       \hline 
       \rule{0pt}{9pt}Brane & $x_0$ & $x_1$ &$x_2$ &$x_3$ & $r$ & $z_1$ & $z_2$ & $w_1$ & $w_2$ &$w_3$  \\\hline \hline
       \rule{0pt}{10.pt}D3  & $\mathsf X$ & & & & $\mathsf X$ & $\mathsf X$ & $\mathsf X$ & & & \\   \hline
       \rule{0pt}{10.pt}D5  & $\mathsf X$ & $\mathsf X$ & $\mathsf X$ & & & & & $\mathsf X$ & $\mathsf X$ & $\mathsf X$\\ \hline
       \rule{0pt}{10.pt}F1  & $\mathsf X$ &  & & $\mathsf X$  & & & & & &\\ \hline
    \end{tabular}
\end{equation}
\normalsize
Here $z_{1,2}$, $w_{1,2,3}$ are local coordinates on $S^2$ and $S^3$, respectively. 
The relevant brane linking
in our system is measured by
the following quantity $L$ defined modulo $M$,
\be
L=\int_{M_2 \times S^3} F_5 =- \int_{M_1 \times S^2} F_3  = \int_{M_2} db_1= -\int_{M_1} dc_0 \,,
\ee
where  $ M_2 = \mathbb R_{x_1} \times   \mathbb R_{x_2}$ and $M_1=\mathbb R_{r}$. 
On the worldvolume of $\cN_3^{(1)}$, the EOM for $a$ implies
$db_1 = - M da$. Thus, $db_1$ is exact modulo $M$. As a result,
the linking $L$ must be conserved
modulo $M$.
When the D3 crosses the D5, this changes to $db_1=-M da + \delta({\rm pt} \subset M_2)$. The localized source is the effect of a new object (an F1-string) that is created, which intersects both $M_2$ and $M_3$ and extends along ${t=x_0,x_3}$, figure \ref{fig:HW}.
The system in  \eqref{eq:branesyst} is related to the original Hanany-Witten setup NS5-D5-D3 by S- and T-dualities. The D3- and D5-branes link in the direction $x_3$, this means that an F1 is created when the the D3 crosses the D5 \eqref{eq:branesyst}. F1 strings are indeed electrically charged under $e^{-2\pi i \oint b_2}$, which was precisely the dressing for $\mathcal{O}_2(M_2)$. 
This also matches the physics of the action on the 't Hooft loop in $PSU(M)$ through a non-invertible domain wall between de-/confining vacua, that mimicks closely the order/disorder transition in the Ising model.


\newpage

\smallskip

\noindent{\bf Outlook.}
We provide a bottom up approach --  via Gauss law constraints in supergravity --  and top down one --  via branes in string theory -- for constructing symmetry operators in holography.  Our methods are crucial for a systematic extraction of symmetry defects, whenever SymTFTs are available. It deserves further study. 
Future applications include theories that have similar type of mixed anomalies in the SymTFT, such as $\mathcal{N}=4$ SYM theories holographically dual to AdS$_5 \times S^5$ with non-invertible duality defects \cite{Choi:2021kmx, Kaidi:2022uux}.
A similar realization of these topological defects in terms of M5-branes at the boundary of conical in $G_2$-holonomy spaces is also tempting, and show similar features to \eqref{eq:branesyst}. 
Finally we also briefly comment on the holographic realization of the (self-) duality and triality of non-invertible topological defects \cite{Choi:2022zal} for $\mathcal{N}=4$ SYM in AdS$_5 \times S^5$. Duality and Triality are all subgroup of $SL(2,\mathbb{Z})$ symmetries, therefore is very tempting to conjecture that the topological defects in this case are engineered by 7-branes wrapping $S^5$. These are just example of possible applications of this approach, which we plan to come back in the future, but very importantly they show the broader scope of the holographic supergravity and brane approach, which are meant to address questions about symmetries of the QFTs living at the boundary.


\bibliography{GenSym}


\onecolumngrid

\vspace{0.2cm}
\noindent
\textbf{Acknowledgements.}
We thank Lakshya Bhardwaj, Lea Bottini, Antoine Bourget, Inaki Garcia-Extebarria, Zohar Komargodski, Carlos Nu\~nez, Ashoke Sen, Thomas Waddleton for discussions.  
FB and SSN are supported by the ERC grant 682608. SSN also acknowledges support through the Simons Foundation Collaboration on ``Special Holonomy in Geometry, Analysis, and Physics", Award ID: 724073, Schafer-Nameki.  IB is supported in part by NSF grant
PHY-2112699 and by the Simons Collaboration on Global Categorical Symmetries.  FA, IB and FB are grateful for the hospitality provided by The Lewiner Institute for Theoretical Physics at the Technion - Israel Institute of Technology.  

\noindent
\textbf{Note.} A related paper \cite{GarciaEtxebarria:2022vzq} will appear at the same time and we thank the author for coordinating submission.


\appendix

\section{Supergravity Analysis and (Non-Invertible) Symmetry Generators}
\label{app_bulk}

The Klebanov-Strassler setup \cite{Klebanov:2000hb} provides a holographic
realization of 4d $\mathcal N = 1$
SYM theory in type IIB string theory. It consists of two main ingredients. First, a stack of $N$ D3-branes, extending along $\mathbb R^{1,3}$ and situated at the tip of
the conifold, \emph{i.e.}~the
non-compact Calabi-Yau metric cone
over $T^{1,1}$, a Sasaki-Einstein
5-manifold with topology $S^2 \times S^3$. The near-horizon geometry at this stage is $AdS_5 \times T^{1,1}$, supported by $N$ units of $F_5$ flux.
The second ingredient is a stack of $M$ D5-branes wrapping the $S^2$ inside $T^{1,1}$ and extending
on $\mathbb R^{1,3}$. The backreaction
of the $M$ D5-branes has two important effects: the $AdS_5$ metric is deformed (see
$\d s^2_{W_5}$ below \eqref{eq:conmet}); the $F_5$ flux
on $S^5$ is no longer constant,
see \eqref{eq:fluxKS}.
This appendix focuses on the low-energy 5d effective action on $W_5$
obtained from reduction of the 10d couplings of type IIB supergravity on the horizon $T^{1,1}$, threaded
by the fluxes in \eqref{eq:fluxKS}.

\paragraph{Bulk topological couplings from
anomaly inflow and consistent truncation.}
The relevant terms in the 5d bulk effective action take the form 
$S = S_{\rm kin} + S_{\rm top}$.
In our conventions, the action
enters the path integral as $e^{ i S}$.
The kinetic terms 
are standard, \emph{i.e.}~a
sum of terms of the form $f_1 \wedge * f_1$
and similar for the other field strengths,
with constant coefficients.
The topological terms $S_{\rm top}$
are  written as an integral over a 6-manifold
$W_6$ such that
$\partial W_6 = W_5$,
\begin{align} \label{top_action}
 S_{\rm top}  =2\pi  \int_{W_6}  \Big[
& N h_3  f_3 
+ \cF_2  g_2^2  
- f_1  g_2  h_3
 - \tfrac 14 N^2 \cF_2^3  
\Big] \ .
\end{align}
These  couplings 
can be derived 
via anomaly inflow in type IIB supergravity. The starting point is
the 11-form \cite{Bah:2020jas}
\be   \label{I11}
\mathcal I_{11} = \tfrac 12 \cF_5 d\cF_5 - \cF_5 H_3 F_3 \ .
\ee 
This object captures both the Chern-Simons coupling 
in the 10d type IIB supergravity action, and the effect of 
the selfduality of $F_5$. The quantity $\cF_5$ is a non-closed
5-form, related to $F_5$ as $F_5 = (1+*_{10})\cF_5$.
The task at hand is to expand $F_3$, $H_3$, $\cF_5$ onto
cohomologically non-trivial forms of the horizon $T^{1,1}$.
The latter is topologically $S^2 \times S^3$. Its Einstein metric
reads $ds^2(T^{1,1}) = \tfrac 49 D\psi^2 
+ \tfrac 16 ds^2(S^2_1) + \tfrac 16 ds^2(S^2_2)$,
where $ds^2(S^2_i) = d\theta_i^2 + \sin^2 \theta_i d\phi_i$
($i=1,2$) are round metric on unit-radius $S^2$'s,
the angle $\psi$ has period $2\pi$, 
and $D\psi = d\psi + \tfrac 12 \cos \theta_1 d\phi_1
+ \tfrac 12 \cos \theta_2 d\phi_2$.
We turn on a Kaluza-Klein 1-form gauge field $A_1$
associated to the isometry $\partial_\psi$,
by means of the replacement $\frac{d\psi}{2\pi} \rightarrow \frac{ d\psi}{2 \pi} + A_1$.
Let us define
$V_i = \frac{1}{4\pi} \sin \theta_i d\theta_i d\phi_i$ ($i=1,2$).
The expansion of $H_3$, $F_3$, $\cF_5$ reads
\be  \label{inflow_expansion}
H_3 = h_3 \ , \qquad F_3 = M \omega_3 + f_1 \omega_2  + f_3
 \ , \qquad
 \cF_5 = N \mathcal V_5 + g_2 \omega_3  \ .
\ee 
The quantities $h_3$, $f_1$, $f_3$, $g_2$ are 
the field strengths of external gauge potentials.
The integers $M$, $N$ capture the D5-
and D3-brane charges in the Klebanov-Strassler setup.
We have introduced $\omega_3 = (V_1 - V_2) \frac{D\psi}{2\pi}$,
$\omega_2 = -\tfrac 12 (V_1 - V_2)$, $\mathcal V_5 = \omega_2 
\omega_3 + \tfrac 12 \cF_2 (V_1 + V_2) \frac{D\psi}{2\pi}$,
where $\cF_2 = dA_1$.
The 3-form $\omega_3$ is dual to the non-trivial 3-cycle
in $T^{1,1}$, which can be represented by the $S^1_\psi$
fibration over $S^2_1$ at a generic point on $S^2_2$.
The 2-form $\omega_2$ is dual to the non-trivial 2-cycle
in $T^{1,1}$ and is normalized by requiring $\int_{T^{1,1}}\omega_2 \omega_3 = 1$, which
encodes the fact that the 2-cycle and 3-cycle in $T^{1,1}$
have intersection number 1.
We have $d\omega_2 = 0$, $d\omega_3 = - 2 \omega_2 
\cF_2$, $d\mathcal V_5 = \frac 12 (V_1 + V_2) \cF_2^2$.
In particular, the definition of $\cV_5$ is engineered in such a way that $d\mathcal V_5$ contains no terms linear in
$\cF_2$. This is necessary to ensure the compatibility of
\eqref{inflow_expansion} with the type IIB Bianchi identities
$dF_5 = H_3 F_3$, $dH_3 = 0$, $dF_3 = 0$ (we have $F_1 \equiv 0$). The 10d Bianchi identities also  imply
the Bianchi identities \eqref{Bianchi} for the external
field strengths.
The 6-form integrand in \eqref{top_action}
can now be readily reproduced
by plugging \eqref{inflow_expansion} into
\eqref{I11} and fiber-integrating along
$T^{1,1}$.

The topological couplings \eqref{top_action}
can also be inferred from the consistent
truncation of \cite{Cassani:2010na}.
With reference to the full set of topological
couplings in the 5d action of 
\cite{Cassani:2010na}
(and using their notation)
we freeze $C_0$ to a constant,
we gauge fix the scalars $b^J$, $c^J$ to zero,
we set to zero the massive 1-form fields $b_1$, $c_1$
and the massive scalar $b^\Phi$,
we set
$p=0$, $q=M$, $k=N$.
The couplings \eqref{top_action} are reproduced 
 with the identifications
$(A_1, g_2,f_1, h_3 , f_3)_\text{here} = (-\frac 12 A, f_2^\Phi, Dc^\phi, db_2, dc_2 )_\text{there}$.
The term $\cF_2^3$ in \eqref{top_action}
originates from 
$dA (d \widetilde a_1^J)^2$ in \cite{Cassani:2010na},
due to the fact that 
$A + \widetilde a_1^J$
is a massive vector. Since we are interested in massless modes, we can set effectively
$\widetilde a_1^J = -A$.

\paragraph{Dualization of $c_0$.}
After solving the Bianchi identities \eqref{Bianchi} in terms of 
base-point fluxes and globally defined
gauge potentials, we can write the topological
action \eqref{top_action} as an integral of a 
globally defined 5-form on $W_5$,
\begin{align}
S_{\rm top} &=2\pi \int_{W_5} \Big[
N b_2 (dc_2 + f_3^{\rm b})
+  A_1 (g_2^{\rm b})^2
+ b_2 g_2^{\rm b} f_1^{\rm b}
+ \tfrac{1}{2 M} \widetilde f_1  
(\widetilde g_2^2  +2 \widetilde g_2 g_2^{\rm b}
- \tfrac 14 N^2 (dA_1)^2)
 + \tfrac{1}{2M} f_1^{\rm b} \widetilde g_2^2
\Big] \ ,
\end{align}
where we have 
introduced the shorthand notation
$\widetilde f_1 = dc_0 + 2M A_1$,
 $\widetilde g_2  = d\bbeta_1 + M b_2$, and 
suppressed wedge products for brevity.
We add a Lagrange multiplier to the  
action, implementing the Bianchi identity for
$\widetilde f_1$, 
\be
S_{\rm mult} =2\pi \int_{W_5} (2MdA_1 - d\widetilde f_1   ) c_3 \ ,
\ee
where $c_3$ is a
globally defined 3-form potential. After integrating
by parts the term $d\widetilde f_1 c_3$, the total
action $S_{\rm kin} + S_{\rm top} + S_{\rm mult}$
depends algebraically on $\widetilde f_1$,
which can be integrated out using its classical
  equation of motion  $ f_1 \propto * f_4$
(recall that $f_1 = f_1^{\rm b} + \widetilde f_1$),
with $f_4$ given as
\be  \label{f4_def}
f_4 = dc_3 - \tfrac{1}{2M} \big( \widetilde g_2^2  +2 \widetilde g_2 g_2^{\rm b}
- \tfrac 14 N^2 (dA_1)^2 \big) \ .
\ee 
The new action after 
eliminating $\widetilde f_1$
has standard kinetic terms
and a new set of topological
couplings as in \eqref{new_top}.
More precisely, to reproduce  
 \eqref{new_top}
we also have to
add  some total derivatives to the 5d action,
constructed with the globally-defined gauge
potentials and the closed base-point fluxes.

\paragraph{Canonical momenta and Gauss constraints.}
We consider a 5d spacetime of the form
$W_5 = M_4 \times \mathbb R_t$ with product
metric $ds^2(W_5) = ds^2(M_4) -dt^2$.
We write the 5d exterior derivative  as
$d = d_4 + dt \partial_t$.
We decompose the gauge potentials
in components without and with a leg along
the time direction $t$, (where we denote the spatial slice part with a bar)
\be
A_1 = \overline A_1 + dt \, A^0_0 \ , \qquad
\bbeta_1 = \overline \bbeta_1 + dt \, \bbeta^0_0 \ , \qquad
b_2 = \overline b_2 + dt \, b^0_1 \ , \qquad
c_2 = \overline c_2 + dt \, c^0_1 \ , \qquad 
c_3 = \overline c_3 + dt \, c^0_2 \ . 
\ee  
The Lagrangian   $\cL$ is defined by
$S =2\pi \int dt \mathcal L$, and it is the integral
over $M_4$ of a 4-form Lagrangian density. 
The canonical momenta associated to the
spatial components $\overline A_1$, $\overline \bbeta_1$,
$\overline b_2$, $\overline c_2$, $\overline c_3$
are defined as variational derivatives
of $\cL$ with respect to $\partial_t \overline A_1$, $\partial_t \overline \bbeta_1$, and so on.
Similarly, the   Gauss constraints
are defined as the variational derivatives of
$\cL$ with respect to the time components
$A^0_0$, $\bbeta^0_0$, and so on,
\begin{align}
\delta \cL  = \int_{M_4} \Big[  
& \Pi_{A_1} \delta \partial_t \overline A_1
+ \Pi_{\bbeta_1} \delta \partial_t \overline \bbeta_1
+ \Pi_{b_2} \delta \partial_t \overline b_2
+ \Pi_{c_2} \delta \partial_t \overline c_2
+ \Pi_{c_3} \delta \partial_t \overline c_3
 + \cG_{A_1} \delta A^0_0
+ \cG_{\bbeta _1} \delta \bbeta^0_0
+  \cG_{b_2} \delta b^0_1
+  \cG_{c_2} \delta c^0_1
+  \cG_{c_3} \delta c^0_2
\Big] \ , \nn
\end{align}
where we have denoted the momenta with $\Pi$
and the Gauss constraints with $\cG$.
We compute
\begin{align}  \label{momenta}
\Pi_{b_2} &= \widetilde \Pi_{b_2} - \tfrac 12 N
\overline c_2 \ , & 
\Pi_{c_2} &= \widetilde \Pi_{c_2} + \tfrac 12 N
\overline b_2 \ , 
&
\Pi_{A_1} &= \widetilde \Pi_{A_1} - M \overline c_3 \ , &
\Pi_{c_3} &= \widetilde \Pi_{c_3} - M \overline A_1 \ , &
\Pi_{\bbeta_1} &= \widetilde \Pi_{\bbeta_1} \ , 
\end{align}
where we used a tilde to denote the contributions
originating from the kinetic terms. 
The Gauss constraints are reported in \eqref{Gauss}.
The quantities $\widetilde \Pi$, $\widetilde \cG$
satisfy 
\be   \label{kinetic_relations}
\widetilde \cG_{A_1} = -d_4 \widetilde \Pi_{A_1} \ ,
\qquad
\widetilde \cG_{\bbeta_1} = -d_4 \widetilde \Pi_{\bbeta_1} \ , 
\qquad
\widetilde \cG_{b_2} = d_4 \widetilde \Pi_{b_2} 
+ M \widetilde \Pi_{\bbeta_1} \  , \qquad
\widetilde \cG_{c_2} = d_4 \widetilde \Pi_{c_2} \ , \qquad
\widetilde \cG_{c_3} = - d_4 \widetilde \Pi_{c_3} \ .
\ee 
This can be seen  by
noting that the part of the 4-form Lagrangian 
density that originates from the 5d kinetic terms
depends on the various gauge potentials only
via the following combinations:
$d_4 \overline A_1$, $d_4 \overline b_2$,
$d_4 \overline \bbeta_1 + M \overline b_2$,
$d_4 \overline c_2$,
$d_4 \overline c_3$
$\partial_t \overline A_1 - d_4 A^0_0$,
$\partial_t \overline b_2 - d_4 b^0_1$,
$\partial_t \overline \bbeta_1 - d_4 \bbeta^0_0 + M  b^0_1$,
$\partial_t \overline c_2 - d_4 c^0_1$,
$\partial_t \overline c_3 - d_4 c^0_2$.
The relations \eqref{kinetic_relations} then follow
from the chain rule for functional derivatives.
Combining \eqref{Gauss}, \eqref{momenta},
and \eqref{kinetic_relations} we arrive
at the expression for the Gauss
constraints in terms of the canonical momenta,
\be   \label{Gauss_with_momenta}
\begin{array}{l}  
\cG_{b_2} =  d_4 \Pi_{b_2} + M \Pi_{\bbeta_1}
- \tfrac 12 N d_4 \overline c_2 - N f_3^{\rm b} \ ,
\\
\cG_{c_2} =  d_4 \Pi_{c_2}  
+ \tfrac 12 N d_4 \overline b_2  \ ,  
\end{array} \qquad
\begin{array}{l}
\cG_{A_1} = - d_4 \Pi_{A_1}  
+ M d_4 \overline c_3  +(g_2^{\rm b})^2 \ ,
\\
\cG_{c_3} =  - d_4 \Pi_{c_3}  
+ M d_4 \overline A_1  \ ,  
\end{array} \qquad
\cG_{\bbeta_1} = d_4 \Pi_{\bbeta_1} \ .
\ee

\paragraph{Symmetry Generators from Gauss law constraints.}\label{ref:SymGL}
We briefly review why the Gauss law constraint can be used to construct symmetry operators.  For this we consider a $p$-form gauge field $A_p$ with transformation $A_p\to A_p + d \lambda_{p-1}$ on a $d+1$-dimensional spacetime with constant time slices $W_d$.  Let $\mathcal{G}_A$ be the Gauss constraint, which is a closed ${(d-p+1)}$-form.  In Hamiltonian quantization, it generates a small gauge transformation as the operator
\begin{equation}
    \exp \left( 2\pi i \int_{W_d} \lambda_{p-1} \wedge \mathcal{G}_A \right). \label{gausslaww}
\end{equation} Now we consider a singular gauge transformation that has support on a boundary surface $M_{d-p} = \partial M_{d-p+1}$ as 
\begin{equation}
    d\lambda_{p-1} = \delta \left(M_{d-p} \right). 
\end{equation} This allows us to associate a symmetry generator for gauge transformation on the pair $(M_{d-p}, M_{d-p+1})$ as
\begin{equation}
    \exp \left( 2\pi i \int_{M_{d-p+1}}  \mathcal{G}_A \right). 
\end{equation}  
If $\mathcal{G}_{A} = d \mathcal{O}_{A}$, with $\mathcal{O}_A$ a globally defined operator, we can integrate by parts to obtain a genuine dynamical operator defined on $d-p$-cycles as $e^{2\pi i \int_{M_{d-p}} \mathcal{O}_A}$. Now as we take the $W_4$ near the boundary of $W_5$, the kinetic terms are suppressed and the operator is topological on $M_{d-p}$ \cite{Witten:1998wy, Belov:2004ht}. 

If $\mathcal{G}_a$ cannot be expressed as a derivative of a gauge invariant and globally defined operator, consider 
$M_{d-p} = \partial M_{d-p+1}$ and define  \be
\mathcal{S}_{A} (M_{d-p})= e^{2\pi i \int_{M_{d-p+1}} \mathcal{G}_A }\,.
\ee  
In this operator the contribution from the kinetic terms are derivatives of local operators. The main task is now to define a genuine operator on $M_{d-p}$. 

Let us concretely carry this out in the case of $A= A_1$ with Gauss law constraints 
\be
\cG_{A_1}  =\widetilde \cG_{A_1}
+2M d_4 \overline c_3
+(\overline g_2^{\rm b})^2 \,.
\ee
Here $\widetilde{\mathcal{G}}_{A_1} = d\widetilde{\mathcal{O}}_{A_1}$ is exact and $\bar{c}_3$ is globally defined.
We wish to define a genuine operator
on $M_3 = \partial M_4$ such that,
when raised to the $2M$-th power
it reproduces
the operator constructed from the Gauss law constraint $\mathcal{S}_{A_1}(M_4)$.  At this stage, we will also consider the operator near the boundary of $W_5$ and therefore can drop the contribution from the kinetic term.  However we must now consider the different choices of boundary conditions.  The first case we fix $b_2$ and thus $\overline g_2^{\rm b}$
at the boundary as classical backgrounds.
This corresponds 
to 4d $\cN = 1$ $SU(M)$ SYM,
with a global electric 
$\mathbb Z_M$ 1-form symmetry
coupled to a non-dynamical
discrete 2-form field.
The genuine operator on $M_3$ in this case
is simply the standard holonomy
$e^{2\pi i \int_{M_3} \overline c_3}$
accompanied by  the 
$c$-number phase 
$e^{2\pi i \frac{1}{2M}\int_{M_4}(\overline g_2^{\rm b})^2}$. This operator
obeys group-like fusion rules.

The more interesting scenario is when  we sum over $\overline g_2^{\rm b}$
at the boundary.
In field theory, we gauge the electric 1-form
symmetry of 4d $\cN =1$ $SU(M)$ SYM,
thereby getting the $PSU(M)$ theory.  Here we cannot treat $e^{2\pi i \frac{1}{2M}\int_{M_4}(\overline g_2^{\rm b})^2}$ as a c-number. However we observe that
\begin{equation}
    e^{2\pi i \int_{M_4}(\overline g_2^{\rm b})^2} = \int \mathcal{D} a \mathcal{D} c e^{-2\pi i \int_{M_4} \left(M^2 da \wedge da + 2M da \wedge \overline g_2^{\rm b} + d c \wedge \overline g_2^{\rm b}\right) }\,,
\end{equation} 
where $\overline g_2^b$ is flat on the left hand side. Therefore integrating over $a$ and $c$ reproduces the right hand side. Now we are free to write 
\begin{align}
    e^{2\pi i \int_{M_4}(\overline g_2^{\rm b})^2} &= \int \mathcal{D} a \mathcal{D} c e^{-2\pi i \int_{M_4} \left(M^2 da \wedge da + 2M da \wedge \overline g_2^{\rm b} + d c \wedge \overline g_2^{\rm b}\right) } = \int \mathcal{D} a ~e^{-2\pi i~ 2M \int_{M_3} \left(\frac{M}2 a \wedge da +  a \wedge \overline g_2^{\rm b} \right)}\,.
\end{align}  
 In the middle term, the integral localizes to configurations that satisfy $d(Mda + g_2^{\rm b} )=0$ and $d g_2^{\rm b} =0$ thereby realizing the first equality.
In the last expression we have integrated over $c$ and integrated by part to $M_3$.  Now we have rewritten the right hand side in a way that we can take the $2M$-th root.
The genuine operator on $M_3$ can then be determined using this prescription. For the anomaly in the solution at hand we find  \eqref{good_operator}, which precisely has 
the non-invertible fusion rule \eqref{eq_fusion}.

\section{D5-brane action and reduction} \label{app:D5brane}
The topological couplings in the D5-brane action
are encoded in the 6-form
\begin{align}
I_6^{\text{D5}(0)} & = \Big[  e^{da  - B_2} (C_6 + C_4 + C_2 + C_0) \Big]_{\text{6-form}}
 = C_6 + C_4 (da  - B_2)
+ \frac 12 C_2 (da  - B_2)^2
+ \frac 16 C_0 (da  - B_2)^3 \ .
\end{align}
We neglect contributions from the tangent and normal bundles.
The associated 7-form anomaly polynomial is
\begin{align}
I_7^{\text{D5}} = dI_6^{\text{D5}(0)} & = 
F_7 + F_5 (da  - B_2)
+ \frac 12 F_3 (da  - B_2)^2
+ \frac 16 F_1 (da  - B_2)^3 \ ,
\end{align}
where
$F_7 = dC_6 - H_3 C_4$,
$F_5 = dC_4 - H_3 C_2$,
$F_3 = dC_2 - H_3 C_0$,
$F_1 = dC_0$.
In these conventions, the Bianchi identities read
$dF_p =  H_3 F_{p-2}$
($p=3,5,7$) and $dF_1  =0$.
To reduce $I_7^{\text{D5}}$ on $S^3$, we use
\be
F_7 = f_4   {\rm vol}_{S^3} + \dots \ , \quad
F_5 = g_2  {\rm vol}_{S^3} + \dots \ , \quad
F_3 = M   {\rm vol}_{S^3} + \dots \ , \quad
F_1 = 0 + \dots \ , \quad
B_2 = b_2 + \dots
\ee
where the ellipses stand for contributions 
that are not relevant for the $S^3$ reduction.
The 10d Bianchi identities require in particular
$df_4 =  h_3 g_2$,
$dg_2 =  M h_3$,
where $h_3 = db_2$.
 We obtain
\begin{align}
I_4^{\text{D5}}  := \int_{S^3} I^{\text{D5}}_7 & = f_4 + g_2 (da  - b_2)
+ \frac 12 M (da  - b_2)^2 \ ,
\end{align}
with $\int_{S^3} {\rm vol}_{S^3} = 1$.
Let us collect powers of the D5-brane gauge field $a$,
\be
I_4^{\text{D5}} = \bigg(f_4 - g_2 b_2  + \frac 12 M b_2^2 \bigg)
+ da  ( g_2  - M b_2   ) + \frac 12 M da  da  \ .
\ee
The Bianchi identity for $g_2$ is solved by setting $g_2 = db_1 + M b_2$, hence $g_2 - M b_2 = db_1$.
(Here we do not separate the base-point for $g_2$ explicitly; it is understood that $b_1$ can be topologically non-trivial.)
The Bianchi identities for $f_4$ and $g_2$ imply that
$
d(f_4 - g_2 b_2  + \frac 12 M b_2^2)
= 0
$. 
We may therefore introduce a bulk 3-form gauge potential
$c_3$ satisfying
\be
f_4 - g_2 b_2  + \frac 12 M b_2^2 = dc_3 \ .
\ee
The 4-form anomaly polynomial $I_4$ then reads
\beq
I_4^{\text{D5}}  = dc_3 + da  db_1 + \frac 12 M da  da  \ , \qquad \text{hence} \qquad 
I_3^{\text{D5}(0)}  = c_3 + \frac M2  a  da   + a  db_1  \ .
\ee
The bulk extended operator of interest is the holonomy
of $I_3^{(0)}$
with charge 1,
$e^{2\pi i \int_{M_3} I_3^{(0)}}$. 

The topological terms for a D7-brane
wrapping $T^{1,1} \cong S^3 \times S^2$ are derived
analogously from the 9-form
anomaly polynomial
$I_9^{\text{D7}}   = 
F_9 + F_7 (da  - B_2)
+ \frac 12 F_5 (da  - B_2)^2
+ \frac{1}{3!} F_3 (da  - B_2)^3
+ \frac{1}{4!}  F_1 (da  - B_2)^3
$.  
We take into account the worldvolume gauge flux
on $S^2$ with the replacement $da  \rightarrow f_2^{S^2} + da$, $\int_{S^2} f_2^{S^2} = 2$. We may use 
$\int_{S^2} (f_2^{S^2} + da - B_2)^p = 2p(da - B_2)^{p-1}$. In total,
\be 
I_4^{\text{D7}} := \int_{T^{1,1}} I_9^{\text{D7}} =
2 I_4^{\text{D5}}   \ .
\ee  
We have not included an $F_5 \propto {\rm vol}_{T^{1,1}}$
contribution because this flux is not quantised
in the UV KS solution, see \eqref{eq:fluxKS}.





\section{Review of Myers Effect and Application to D5-branes on $S^3$} \label{app:Myers}
In this appendix we briefly review the main aspects of the Myers effect \cite{Myers:1999ps}, applying it to $k$ D5-branes wrapping $S^3$ in the Klebanov-Strassler geometry. Before proceeding with the details of the explicit construction that we consider, let us set the basis by providing the general expression for the non-Abelian DBI and WZ action of the brane. First of all the coordinate orthogonal to the D$p$-brane stack are promoted to matrices in the adjoint of  $\mathrm{U}(k)$, $X^i$. The action is then
\begin{equation}
\label{eq:nabAction}
S^{\rm non-Abelian}_{\mathrm{D}p}\,=\,-\mu_p\,\int d^{p+1}x\,\text{STr}\left (\,e^{-\phi}\sqrt{-\det g^{\shortparallel}}\sqrt{\det \Theta}-P\left[e^{\frac{1}{2\pi}\iota_X\iota_X}\left(e^{-B}\wedge C\right)\right]\right)\,,
\end{equation}
where $g^{\shortparallel}$ is the pullback of the metric onto the branes, $\text{STr}$ is the symmetrized trace and $P$ the pullback on the worldvolume. The matrix $\Theta$ is 
\begin{equation}
\Theta^i{}_j\,=\,\delta^i{}_j+\frac{1}{2\pi}\left[X^i,X^k\right]\left( g^{\bot}_{kj}-B_{kj} \right)\,,
\end{equation}
where $g^{\bot}$ is metric perpendicular to the branes. In the WZ action we have the contraction $\iota_X$, which is given by the action of the operator $X^i\partial_{x^i}$ on the potentials. For instance, on a two-form $C^{(2)}=\frac{1}{2}C_{ij} dx^i\wedge dx^j$:
\begin{equation}
\iota_X\iota_X C^{(2)}\,=\,X^jX^i\,C_{ij}\,=\,\frac{1}{2}C_{ij}\left[X^j,X^i\right]\,.
\end{equation}
To perform explicit computation with the action \eqref{eq:nabAction} we need to Taylor expand up to two-derivatives terms and at most quartic scalar interactions, which can be thought as first order correction given by the non-Abelian action with respect to the Abelian one. 

Let us go back to the setup we are studying. The metric of the KS solution for large values of the radial coordinate is given by the Klebanov-Tseytlin metric \cite{Klebanov:2000nc},
\be\ba \label{eq:conmet}
\d s^2_{10} &=\d s^2_{W_5} + \cR^2(r)\d s^2_{T^{1,1}}\,, \quad \cR(r) \sim \text{ln}\left({r /r_s}\right)^{1/4}\,,
\ea\ee
where $\d s^2_{W_5}=\frac{r^2  \d \vec{x}^2}{\cR^2(r)}+{\cR^2(r) \d r^2  \over r^2} $, and  $r_s =r_0 e^{ -\frac{2 \pi N}{3 g_s M^2} -\frac{1}{4}}$. 
The non-trivial fluxes in the background are 
\be \label{eq:fluxKS}
{\int_{S^{3}}} F_{3}  = M\,, \quad  
{\int_{S^{2}}} B_{2}  = \cL(r) \,, \quad 
{\int_{T^{1,1}}}F_{5} =\cK(r) = N+M \cL\,, \quad 
\mathcal{L} := \frac{3g_{s}M}{2\pi} \text{ln}(r/r_{0})\,.
\ee
The stack of $k$ D5-branes wraps $S^3$ engineering the topological defects, and we would like to study its dynamics close to the boundary at $r\rightarrow r_0$, with $r_0>>0$, in order for the supergravity regime to be reliable. The orthogonal directions are spanned by $\{x_3,r,\theta, \gamma\}$, where the last two are the coordinates of $S^2\subset T^{1,1}$. Let us put aside $x_3$, which will not affect the analysis of this section. Since we are in a regime where for $r\rightarrow \infty$ the radial dynamics of the entire $k$ D5-branes system is frozen and we focus on the dynamics of the relative positions of the $k$ D5-branes, we also fix the warping factor $\mathcal{R}(r=r_0)$. The geometry spanned by the coordinates $\{r,\theta, \gamma\}$ orthogonal to the D5 brane stack in the limit $r\rightarrow r_0$ is $\mathbb R^3$ where the metric is $ds^2_{\mathbb{R}^3_{\{r,\theta,\gamma\}}}= \frac{1}{2r_0^2}(dr^2 +r^2 d\Omega_2)$. The $B$ components in  $\mathbb R^3$ which depend on $X^i$ are then given by
\begin{equation}
    B_{ij}= \frac{1}{r_0^3}\epsilon_{ijk}X^k.
\end{equation}
The fluxes \eqref{eq:fluxKS} allow for no WZ action for the D5 stack wrapping $S^3$ since the only background which pulls back to the brane is $B_{2}$. Hence we need to expand the DBI action, which up to quadratic order reads,
\begin{equation}
\label{eq:nabDBI}
S^{\rm non-Abelian}_{\mathrm{D}5}\,=\,-\mu_5\,\text{Tr}\int d^{6}x\sqrt{g^{\shortparallel}}\left(1+\frac{1}{2} g^{\mu \nu}_{\shortparallel}g^\bot_{ij}\partial_\mu X^i \partial_\nu X^j -\frac{1}{4\pi}B_{ij}[X^j, X^i]+\frac{1}{16\pi^2}[X^i, X^j]g^\bot_{jk}[X^k, X^l]g^\bot_{li}\right)\,.
\end{equation}
The static solutions are described by
\begin{equation}
    [[X^i,X^j],X^j]-2[X^k,X^j]\epsilon_{kj}{}^i=0 \,,
\end{equation}
where we rescaled $X^i \rightarrow r_0 X^i$.
Commuting matrices are always solutions, however the minimal energy solution is one where the fiels are non-commutative ones such that $X^i\sim \alpha^i$, where $\alpha^i$ are the realization as $k\times k$ matrices of $\mathfrak{su}_2$ in the $k$-dimensional representation. The dynamical consequence for the D5 branes is that they puff up and polarize into a D7 with $k$ unit of $f_2^{S^2}$ flux, or alternatively a bound state of D5/D7,  in analogy with the D0/D2 bound state system studied in the original paper \cite{Myers:1999ps}.


\end{document}